\documentclass[12pt,english]{article}
\usepackage[T1]{fontenc}
\usepackage[latin1]{inputenc}

\makeatletter

\makeatletter

\makeatletter

\makeatletter

\AtBeginDocument{\DeclareFontEncoding{T2A}{}{}}

\makeatletter

\AtBeginDocument{\DeclareFontEncoding{T2A}{}{}}

\makeatletter

\AtBeginDocument{\DeclareFontEncoding{T2A}{}{}}

\makeatletter

\makeatletter

\makeatletter

\newtheorem{theorem}{Theorem}

\newtheorem{lemma}{Lemma}

\date{}

\makeatother

\usepackage{babel}
\begin{document}

\title{Phase transitions in one-dimensional static Coulomb media}

\author{Malyshev V. A. \thanks{Faculty of Mechanics and Mathematics, Moscow State University}}
\maketitle
\begin{abstract}
We consider configurations of $N$ charged particles on the interval
with nearest neighbour Coulomb interaction and constant external force.
For different values of external force we find 4 different phases
of the asymptotic particle density for the configuration corresponding
to the minimum of the energy. 
\end{abstract}

\section{Introduction}

The problem of finding $N$ point particle configurations on a manifold
having minimal energy (or even fixed configurations) was claimed important
already long ago \cite{berkenbusch}. That is why we shall say shortly
about the history of this question. We consider systems of particles
with equal charges and Coulomb interaction. Immediately the problem
is separated into two cases: when $N$ is small, where one should
find such configurations explicitely, and the case of large $N$,
where the asymptotics is of main interest. Already J. J. Thomson (discovering
electron in 1897) suggested the problem of finding such configurations
on the sphere, and the answer has been known for $N=2,3,4$ for more
than 100 years, but for $N=5$ the solution was obtained only quite
recently \cite{schwartz}. In one-dimensional case T, J, Stieltjss
studied the problem with logarithmic interaction and found its connection
with zeros of orthogonal polynomials on the corresponding interval,
see \cite{chaitanya}, \cite{ismail}. However, the problem of finding
minimal energy configurations on two-dimensional sphere for any $N$
and power interaction (sometimes it is called the seventh problem
of S. Smale, it is also connected with the names of F. Risz and M.
Fekete) was completely solved only for quadratic interaction (see
\cite{smale}, \cite{dimitrov}, \cite{kuijlaars} and review \cite{nerattini}).
For more general compacts see review \cite{korevaar}.

Here we follow alternative direction: namely, we study how the configuration
could change in the presence of weak or strong external force. It
appears that even in the simplified one-dimensional model with nearest
neighbour interaction there is an interesting structure of fixed points
(more exactly, fixed configurations), rich both in the number and
in the charge distribution. For the constant force case we find 4
phases of the charge density, with respect to the parameter - the
ratio of the constant of interaction strength and the value of external
force. 

We call Coulomb media the space of configurations 
\[
-L\leq x_{N}<...<x_{1}<x_{0}\leq0
\]
of $N+1$ point particles with equal charges on the segment $[-L,0]$.
Here $N$ is assumed to be sufficiently large, however some results
are valid for any $N\geq2$. We assume repulsive Coulomb interaction
of nearest neighbours, and external force $\alpha_{ext}F_{0}(x)$,
that is the potential energy is

\begin{equation}
U=\sum_{i=1}^{N}V(x_{i-1}-x_{i})-\sum_{i=0}^{N}\int_{-L}^{x_{i}}\alpha_{ext}F_{0}(x)dx,V(x)=\frac{\alpha_{int}}{|x|}\label{energy_U}
\end{equation}
where $\alpha_{ext},\alpha_{int}$ are positive constants. This defines
the dynamics of the system of charges, if one defines exactly what
occurs with particles $0$ and $N$ in the points $0$ and $-L$ correspondingly.
Namely, we assume completely inelastic boundary conditions. More exactly,
when particle $x_{0}(t)$ at time $t$ reaches point $0$, having
some velocity $v_{0}(t-0)\geq0$, then its velocity $v_{0}(t)$ immediately
becomes zero, and the particle itself stays at point $0$ until the
force acting on it (which varies accordingly to the motion of other
particles) becomes negative. Similarly for the particle $x_{N}(t)$
at point $-L$. 

To discover phase transitions it is common to consider asymptotics
$N\to\infty$, with the parameters $L,l,F_{0}(x)$ being fixed. Then
the fixed points will depend only on the ``renormalized force''
$F=\frac{\alpha_{ext}}{\alpha_{int}}F_{o}$, and we assume that the
renormalized constant $\alpha_{ren}=\frac{\alpha_{ext}}{\alpha_{int}}$
can tend to infinity together with $N$, namely as $\alpha_{ren}=cN^{\gamma}$,
where $c,\gamma>0$. It is eviodent that if $F_{0}\equiv0$, then
for all $k=1,...,N$
\begin{equation}
\delta_{k}=x_{k-1}-x_{k}=\frac{L}{N}\label{F_ravno_0}
\end{equation}
The case when $\alpha_{ren}$ does not depend on $N$ was discussed
in detail in \cite{Mal-1}, there are no phase transitions but it
is discovered that the structure of the fixed configuration differs
from (\ref{F_ravno_0}) only on the sub-micro-scale of the order $N^{-2}$. 

The necessity to consider cases when $\alpha_{ren}$ depends on $N$,
issues from concrete examples where $\alpha_{ren}\gg N$. E.g. the
linear density of electrons in some conductors, see \cite{Ashcroft},
is of the order $N\approx10^{9}m^{-1}$, $\alpha_{int}=\frac{e^{2}}{\epsilon_{0}}\approx10^{-28}$
and $\alpha_{_{ext}}=220\frac{volt}{meter}e=220\times10^{-19}$ (in
SI system). Thus $\alpha_{ren}$ has the order $10^{11}$. This is
close to the critical point of our model, which, as will shown, is
asymptotically $c_{cr}N$.

We study the density $\rho(x)$ (proving its existence), defined so
that for any subintervals $I\subset[-L,0]$ there exist the limits
\[
\rho(I)=\int_{I}\rho(x)dx=\lim_{N\to\infty}\frac{\#\{i:x_{i}\in I\}}{N}
\]
We find four phases: 1) uniform (constant) density, 2) nonuniform
but positive smooth density, 3) continuous density, zero on some subinterval,
4) density of $\delta$-function type. 

One-dimensional case shows what can be expected in multi-dimensional
case, which is more complicated but has great interest in connection
to the static charge distrubution in the atmosphere or in the live
organism. For example case 4) of the theorem 2 is related to the discharge
possibility, as after disappearance of large external force, the big
concentration of charged particles can produce strong discharge.

\subsubsection*{Main results}

\begin{lemma}

Assume that $F_{0}(x)$ is continuous, nonnegative and does not increase,
that is $F(x)\leq F(y)$ if $x>y$. Then for any $N,L,\alpha_{ren}$
the fixed point exists and is unique. If $y$ is such that $F(x)=0,x\geq y,$
and $F(x)>0,x<y,$ then $\delta_{k+1}>\delta_{k}$, if $x_{k+1}<y$. 

\end{lemma}

Further on we assume for simplicity that $F_{0}>0$ is uniform (constant
in $x$).

\begin{theorem} (critical force)

For any $N,L$ there exists $F_{cr}$ such that for the fixed point
the following holds: $x_{N}>-L$ for $F>F_{cr}$ and $x_{N}=-L$ for
$F\leq F_{cr}$. If $F=cN^{\gamma},\gamma>1,$ then for any $c>0$
we have $x_{N}\to0$. If $F=cN$ then $F_{cr}\sim_{N\to\infty}c_{cr}N$,
where 
\begin{equation}
c_{cr}=\frac{4}{L^{2}}\label{c_critical}
\end{equation}

\end{theorem}

\begin{theorem} (four phases)
\begin{enumerate}
\item If $F=o(N)$, then the density exists and is strictly uniform, that
is for all $k=1,...,N$ as $N\to\infty$ 
\begin{equation}
\max_{k}|(x_{k-1}-x_{k})-\frac{L}{N}|=o(\frac{1}{N})\label{th_2}
\end{equation}

\item If $F=cN$ and $0<c\leq c_{cr}$, then $x_{N}=-L$ and the density
of particles exists, is nowhere zero, but is not uniform (not constant
in $x$);
\item If $F=cN$ and $c>c_{cr}$, then as $N\to\infty$ 
\begin{equation}
-L<x_{N}\to-\frac{2}{\sqrt{c}}\label{c_crit}
\end{equation}
and the density on the interval $(-\frac{2}{\sqrt{c}},0)$ is not
uniform;
\item If $F=cN^{\gamma},\gamma>1,$ then the density $\rho(x)\to\delta(x)$
in the sense of distributions.
\end{enumerate}
\end{theorem}

\subsubsection*{Uniqueness - proof of lemma 1}

Put
\[
f_{k}=\delta_{k}^{-2},k=1,;..,N.
\]
At least one fixed point exists because the minimum of $U$ evidently
exists. Any fixed point satisfies the following conditions 
\[
x_{0}=0
\]
-
\begin{equation}
f_{k+1}+F(x_{k})=f_{k},k=1,...,N-1\label{fixed_point_cond_k}
\end{equation}
However, for tha particle $N$ there are two possibilities:
\begin{equation}
f_{N}\geq F(x_{N})\label{fixed_point_cond_N_more}
\end{equation}
if $x_{N}=-L$, and 
\begin{equation}
f_{N}=F(x_{N})\label{f_equals_F}
\end{equation}
if $x_{N}>-L$. 

Forgetting for a while about fixed points, we will consider equations
(\ref{fixed_point_cond_k}) as the equations uniquely defining (by
induction in $k$) the functions $f_{k}$ of $\delta_{1}$, and thus
$\delta_{k}=\frac{1}{\sqrt{f_{k}}}$ and also $x_{k}=-(\delta_{1}+...+\delta_{k})$.
It is evident that $f_{k}$ and $x_{k}$ are decreasing, and $\delta_{k}$
are increasing functions of $\delta_{1}$. Moreover, if $\delta_{1}\to0$
then all $f_{k}\to\infty$, and $\delta_{k}$ and $x_{k}$ tend to
$0$, then for $\delta_{1}$ sufficiently small the inequality (\ref{fixed_point_cond_N_more})
holds. Thus, if $\delta_{1}$ increases, two cases are possible: 1)
there exists $\delta_{1,final}$ such that 
\[
F(x_{N})=f_{N},x_{N}>-L,
\]
At the same time if $\delta_{1}>\delta_{1,final}$ then $F(x_{N})$
and $\delta_{N}$ increase as functions $\delta_{1}$, and $f_{N}$
decreases. that is why $F(x_{N})>f_{N}$. It follows that in this
case there are no other fixed points; 2) such $\delta_{1}$ does not
exist, but then for some $\delta_{1}$ we have
\[
x_{N}=-L,F(x_{N})\leq f_{N}
\]
This defines the unique fixed point.

\subsubsection*{Note about nonuniqueness}

The monotonicity assumption in the uniqueness lemma is very essential.
One can give an example of nonuniqueness, for a function $F_{0}(x)$
with the only maximum, where the number of fixed points is of the
order of $N$ or more. Namely, on the interval $[-1,1]$ put for $b>a>0$
\[
F_{0}(x)=a-2ax,x\geq0
\]
\[
F_{0}(x)=a+2bx,x\leq0
\]
Then there exists $C_{cr}>0$ such that for all sufficiently large
$N$ and $\alpha_{ren}=cN,c>C_{cr}$, one can show using similar techniques
that for any odd $N_{1}<N$ there exists fixed point such that
\[
-1=x_{N}<...x_{N_{1}}<0<x_{N_{1}-1}<...<x_{\frac{N_{1}+!}{2}}=\frac{1}{2}<...<x_{0}<1
\]
Moreover, any such point will give local minimum of the energy.

\subsubsection*{Critical force - proof of theorems 1 and 2.4}

In case of constant positive force it follows from (\ref{fixed_point_cond_k})
that 
\begin{equation}
f_{i}>f_{i+1}\Longleftrightarrow\delta_{i}<\delta_{i+1},i=1,...,N-1,\label{increase-1}
\end{equation}
that is the lengths $\delta_{i}$ of intervals strictly increase with
$i$. That is why 
\begin{equation}
\delta_{1}<\frac{L}{N}\label{delta_1_less}
\end{equation}
Summation (\ref{fixed_point_cond_k}) over $i=1,...,k-1$ gives that
for any $k=1,...,N$,
\begin{equation}
f_{k}=f_{1}-(k-1)F,k=1,...,N\label{f_k_1-1}
\end{equation}
Similarly to (\ref{f_k_1-1}), summing over $i=N-1,...k-1,$ we get
\begin{equation}
f_{k}=f_{N}+(N-k)F\label{f_k_N-1}
\end{equation}
Then from (\ref{f_k_1-1}) we get
\begin{equation}
\delta_{k}=(\delta_{1}^{-2}-(k-1)F)^{-\frac{1}{2}}=\delta_{1}(1-\delta_{1}^{2}(k-1)F)^{-\frac{1}{2}}\label{delta_k-1}
\end{equation}
and (as the fixed point exists) 
\begin{equation}
1-\delta_{1}^{2}(k-1)F>0\label{bound_1}
\end{equation}
or 
\begin{equation}
\delta_{1}<(\frac{1}{(N-1)F})^{\frac{1}{2}}\label{delta_1_2-1}
\end{equation}

To prove theorem 1 consider a simpler auxiliary model with $L=\infty$.
That is we assume that the particles are situated on the interval
$(-\infty,0]$, and the force $F$ is constant on all $(-\infty,0]$.
In this model for any $F>0$ there is unique fixed point, given explicitely
\[
f_{N}=F,f_{k}=(N-k+1)F,k=N-1,...,1
\]
which follows from (\ref{f_k_N-1}). From this we get 
\begin{equation}
\delta_{N}=F^{-\frac{1}{2}},\delta_{k}=\frac{1}{\sqrt{(N-k+1)F}}\label{delta_k_simpler-1}
\end{equation}
and 
\[
-x_{N}=\sum_{k=1}^{N}\delta_{k}=\frac{1}{\sqrt{F}}\sum_{k=1}^{N}\frac{1}{\sqrt{k}}
\]
The relation of this model to the initial is quite simple. If $S\leq L$,
then the fixed ploints for both models coincide. If $S\geq L$, then
$x_{N}=-L$. In fact, assuming that for the critical point in the
main model $x_{N}>-L$, we get contradiction with the auxiliary model.
That is why the critical force can be found from the condition that
$x_{N}=-L$ in the auxiliary model, that is 
\[
F=F_{cr}=(\frac{1}{L}\sum_{k=1}^{N}\frac{1}{\sqrt{k}})^{2}\sim_{N\to\infty}(\frac{2}{L})^{2}N
\]
One can also say that for any $x<0$ there exists unique $F=F_{x}$
such that $x_{N}=x$. 

For $F=cN,c>c_{cr},$ we have 
\[
-x_{N}=L=\sum_{k=1}^{N}\delta_{k}=\frac{1}{\sqrt{F}}\sum_{k=1}^{N}\frac{1}{\sqrt{k}}\sim\frac{2}{\sqrt{c}}
\]
from where (\ref{c_critical}) follows, this gives Theorem 1. Similarly
the theorem 2.4 follows as $x_{N}\to0$ if $F=cN^{\gamma},\gamma>1$.

\subsubsection*{Nonuniform density - proof of theorems 2.2 and 2.3}

Firstly, consider the case $F=cN,c>c_{cr}$. Then for $k=aN$ we have
from (\ref{delta_k_simpler-1}) 
\[
\delta_{k}=\frac{1}{\sqrt{(N-k+1)F}}\sim\frac{1}{\sqrt{(1-a)c}}\frac{1}{N}
\]
That is why the density exists, and moreover it equals zero on $[-L,x_{N}]$
and is nonuniform $[x_{N},0]$. 

Let now $c\leq c_{cr}$. One can assume $\delta_{1}=b\frac{L}{N},0<b=b(N)\leq1$.
Then for $k=aN,a<1,$ we get from (\ref{delta_k-1}) 
\[
\delta_{k}=b\frac{L}{N}(1-L^{2}cb^{2}\frac{k-1}{N})^{-\frac{1}{2}}\sim b\frac{L}{N}(1-b^{2}cL^{2}a)^{-\frac{1}{2}}
\]
Thus, the density is not uniform.

\subsubsection*{Uniform density - proof of theorem 2.1}

From (\ref{delta_1_less}) we have

\begin{equation}
\delta_{1}^{2}(k-1)F\leq\delta_{1}^{2}(N-1)F=o(1)\label{delta_1_cube}
\end{equation}
Then
\[
L=\sum_{k=1}^{N}\delta_{k}=\delta_{1}\sum_{k=1}^{N}(1-\delta_{1}^{2}(k-1)F)^{-\frac{1}{2}}=\delta_{1}\sum_{k=1}^{N}(1+\frac{1}{2}\delta_{1}^{2}(k-1)F+O((\delta_{1}^{2}(k-1)F)^{2})=
\]
\begin{equation}
=N\delta_{1}+\frac{1}{4}\delta_{1}^{3}FN^{2}+o(\delta_{1}^{3}FN^{2})\label{L_sum_1-1}
\end{equation}
But by (\ref{delta_1_cube}) we have
\[
\delta_{1}^{3}FN^{2}=o(N\delta_{1})
\]
and that is why
\[
\delta_{1}=\frac{L}{N}+o(\frac{L}{N})
\]
The result for all $k$ follows from (\ref{delta_k-1}).

\end{document}